\documentclass[12pt]{iopart}

\def\p{\partial}
\def\Re{{\rm Re}}
\usepackage{amssymb,graphicx}

\begin{document}

\title[Discrete boundary treatment for the shifted wave equation
\ldots]{Discrete boundary treatment for the shifted wave equation in
second order form and related problems} 

\author{Gioel Calabrese and Carsten Gundlach}

\address{School of Mathematics, University of Southampton,
Southampton, SO17 1BJ, UK}

\date{\today}

\begin{abstract}

We present strongly stable semi-discrete finite difference
approximations to the quarter space problem ($x>0$, $t>0$) for the
first order in time, second order in space wave equation with a shift
term.  We consider space-like (pure outflow) and time-like boundaries,
with either second or fourth order accuracy. These discrete boundary
conditions suggest a general prescription for boundary conditions
in finite difference codes approximating first order in time, second
order in space hyperbolic problems, such as those that appear in
numerical relativity. As an example we construct
boundary conditions for the Nagy-Ortiz-Reula formulation of the
Einstein equations coupled to a scalar field in spherical symmetry.

\end{abstract}




\section{Introduction}

One of the major obstacles to obtaining long-term stability in
numerical simulations of strongly gravitating systems, such as a binary
black hole space-time, is the proper treatment of boundaries.  In
general, when hyperbolic formulations of Einstein's equations based on
space-like hypersurfaces are used, one can distinguish between two
types of boundaries: inner and outer boundaries.  Whereas the inner
boundary is a space-like hypersurface introduced to excise the
singularity from the computational domain, the outer boundary is an
artificial time-like surface introduced because of limited
computational resources.

If maximally dissipative boundary conditions are used with
symmetrizable hyperbolic fully first order formulations of the
Einstein equations \cite{FM,FR,AY,KST,ST} it is known how to construct
stable schemes to high order of accuracy.  (Another desirable property
of boundary conditions in general relativity is that they are
compatible with the constraints, but we do not consider this here.)
With first order in time and second order in space formulations such
as \cite{SN,BShap}, on the other hand, although the continuum problem
is reasonably well understood \cite{GG1,GG2,BS,reduct}, much less is
known about discretisations.  The issue of constructing stable finite
difference approximations of boundary conditions for first order in
time and second order in space formulations of Einstein's equations
has not yet been investigated and it is the focus of this work.  As a
first step we consider the shifted wave equation in one space
dimension and look at both second and fourth-order accuracy. A
one-dimensional toy problem already captures many of the technical
difficulties that arise in the higher dimensional case.  For recent
progress regarding fully first order or fully second order
formulations, see \cite{SKW,LSU}.

In Section~\ref{Sec:continuum} we review the continuum
initial-boundary value problem for the shifted wave equation, both to
fix the notation and to state the energy estimates that establish
well-posedness of the continuum problem.  Section~\ref{Sec:Summary}
states our main results: a prescription for strongly stable finite
differencing schemes for the quarter space problem ($x>0$, $t>0$),
with second and fourth order accuracy, and for the two cases where the
boundary is time-like or space-like (outflow).  

The main part of the paper is the proof of the stability and accuracy
of these prescriptions at the semi-discrete level.  Instead of
using an energy method, we use the Laplace transform method, as
described in Chapter~12 of \cite{GKO}.  In its final step our
proof relies on plotting a function of a complex variable to show that
it is bounded.  In Section~\ref{Sec:Laplace} we apply this method to
the semi-discrete initial-boundary value problem for the shifted wave
equation and prove strong stability, estimate
(\ref{Eq:strongstabestimate}), and convergence. Finally, we turn the
semi-discrete scheme into a fully discrete one using the fourth-order
Runge-Kutta method for integrating in time. In
Section~\ref{Sec:Numerical} we present numerical tests confirming the
desired order of convergence of the resulting schemes for the wave
equation.  As an application, in Section~\ref{Sec:NOR} we
consider the Nagy-Ortiz-Reula formulation of the Einstein equations in
spherical symmetry, where the boundary conditions introduced in
Section~\ref{Sec:Summary} are generalized to constraint-preserving
boundary conditions.

For the second-order in space wave equation without a shift term,
stable and accurate boundary conditions can be constructed using the
{\em summation by parts rule} \cite{MatNor,Strand}. The proof of
stability and accuracy of these methods uses the existence of a
discrete conserved energy which is the precise analogue of the
continuum energy. In \ref{Sec:DEM} we attempt to generalise these
methods to the wave equation with a shift, but we fail. The reason is
that in the presence of a shift, three separate summation by parts
properties must be obeyed instead of one with vanishing shift, and
this appears to overspecify the finite differencing scheme. This
suggests to us that standard discrete energy methods are not
suitable for the second-order in space wave equation with a shift, and
by extension for other second-order in space hyperbolic systems.

For completeness, and for comparison with the wave equation, we give
the analogous results for the advection equation in
\ref{Sec:advection}. In \ref{Sec:Direct} we briefly compare our
results with those of \cite{SKW}.

\section{The continuum initial-boundary value problem}
\label{Sec:continuum}

The one-dimensional shifted wave equation consists of a system
of coupled linear partial differential equations (PDEs) of the form
\begin{eqnarray}
\p_t \phi &=& \beta \p_x \phi + \Pi + F^{\phi}, \label{Eq:shifted1}\\
\p_t \Pi &=& \beta \p_x \Pi + \p^2_x \phi + F^{\Pi}, \label{Eq:shifted2}
\end{eqnarray}
where $F^{\phi}(x,t)$ and $F^{\Pi}(x,t)$ are forcing terms and the
parameter $\beta$, the {\em shift}, is assumed to be constant.  The
homogeneous version of system (\ref{Eq:shifted1}-\ref{Eq:shifted2}) is
obtained from the one-dimensional wave equation, $\p_{\tilde t}^2 \phi
= \p_{\tilde x}^2 \phi$, after a Galilean change of coordinates, $t =
\tilde t$, $x = \tilde x-\beta \tilde t$ and the introduction of a new
variable $\Pi = \p_t \phi - \beta \p_x \phi$.  (See \cite{C1} and
\ref{Sec:Direct} for why this is better than using $\p_t\phi$ in
finite difference schemes.)  At time $t=0$ we prescribe initial data
\begin{eqnarray}
\phi(x,0) &=& f^{\phi}(x),\\
\Pi(x,0) &=& f^{\Pi}(x).
\end{eqnarray}
Introducing the vector valued function $u(x,t) =
(\phi(x,t),\Pi(x,t),\p_x\phi(x,t))^{\rm T}$, with periodic boundary
conditions on an interval ${\cal D}$ the energy
\begin{equation}
\| u(\cdot,t)\|^2 \equiv \int_{\cal D} [ \phi^2 + \Pi^2 + (\p_x\phi)^2 ] \, \rmd x
\end{equation}
satisfies the estimate 
\begin{equation}
\|u(\cdot,t)\|^2 \le K(t) \left(\| u(\cdot,0)\|^2  + \int_0^t
\|F(\cdot,\tau)\|^2 \rmd\tau\right),
\end{equation}
where $F(x,t) = (F^{\phi}(x,t),F^{\Pi}(x,t),\p_xF^{\phi}(x,t))^{\rm T}$ and
$K(t)$ is a function which is bounded on any finite time interval and
does not depend on the initial data.

We now introduce a boundary at $x=0$ and consider two ``quarter
space'' problems (meaning $x>0$ and $t>0$): one for the pure outflow
case ($\beta \ge 1$) and one for the time-like case ($|\beta|<1$).
No boundary condition is needed in the outflow case. In the
time-like case we impose the Sommerfeld boundary condition
\begin{eqnarray}
w_-(0,t) =  g(t), \label{Eq:Sommerfeld}
\end{eqnarray}
where $w_{\pm} \equiv \Pi \pm \p_x \phi$ are the characteristic
variables and $g$ is a freely specifiable function compatible with the
initial data.  To obtain an energy estimate we take a time derivative
of the energy and use integration by parts to obtain the inequality
\begin{equation}
\fl\frac{\rmd}{\rmd t}\|u(\cdot,t)\|^2 \le -[ \beta\Pi^2 + 2 \Pi\p_x \phi +
  \beta(\p_x\phi)^2 ]_{x=0}+{\rm const} \left(\|u(\cdot,t)\|^2 + \|F(\cdot,t)\|^2\right).
\end{equation}
Rewriting the boundary term as
\begin{equation}
\label{contoutflowbound}
\frac{1}{2} [ (1-\beta) w_-^2 - (1+\beta)w_+^2]_{x=0},
\end{equation}
which is negative definite in the outflow case and bounded by
$\frac{1}{2}(1-\beta) g^2$ in the time-like case, and integrating,
gives the estimate showing strong well-posedness
\begin{equation}
\|u(\cdot,t)\|^2 \le K(t) \left(\| f\|^2  + \int_0^t
(\|F(\cdot,\tau)\|^2  + \delta | g(\tau)|^2 )\rmd\tau\right),
\end{equation}
where $f(x)=u(x,0)$ and $\delta =0$ for the outflow case and
$\delta=1$ in the time-like case.

Using the energy method we have proved the well-known fact that the
initial-boundary value problem for the shifted wave equation is
well-posed in the outflow case with no boundary condition, and in the
time-like boundary case with a Sommerfeld boundary condition. In the
remainder of this paper we investigate the stability of finite
difference discretisations of these two problems. We shall restrict
our discussions to discretising the equations in space but not in time
(the method of lines). This transforms the partial differential
equation into a large coupled system of ordinary differential
equations (the semi-discrete problem) which can be solved by a
standard ODE integrator, such as fourth order Runge-Kutta.

\section{The semi-discrete initial-boundary value problem: summary of results}
\label{Sec:Summary}

In this Section we summarize our results regarding strong
stability and convergence of schemes approximating the quarter space
problem for the shifted wave equation.  We introduce the grid $x_j = jh$, with $j=0,1,2,\ldots$, and the grid
functions $\phi_j(t)$ and $\Pi_j(t)$ approximating the continuum
variables.  In the interior, $j\ge p/2$ (with $p=2,4$ depending on the
accuracy of the scheme), we use the standard centered minimal width
discretization
\begin{eqnarray}
\frac{\rmd}{\rmd t}\phi_j &=& \beta D^{(1)} \phi_j + \Pi_j +F^{\phi}_j\label{Eq:dshifted1gen},\\
\frac{\rmd}{\rmd t} \Pi_j &=& \beta D^{(1)} \Pi_j + D^{(2)} \phi_j +F^{\Pi}_j\label{Eq:dshifted2gen},
\end{eqnarray}
where $D^{(1)}$ and $D^{(2)}$ approximate the first and second
derivatives, respectively.  In the second order accurate case, these
operators are
\begin{equation}
D^{(1)} = D_0,\qquad D^{(2)} = D_+D_-,
\end{equation}
where $D_+u_j = (u_{j+1}-u_j)/h$, $D_-u_j = (u_j- u_{j-1})/h$ and $D_0
= (D_++D_-)/2$.
In the fourth order accurate case we use
\begin{eqnarray}
D^{(1)} &=& D_0\left(1-\frac{h^2}{6}D_+D_-\right),\label{Eq:D1}\\
D^{(2)} &=&
D_+D_-\left( 1 - \frac{h^2}{12}D_+D_-\right).\label{Eq:D2}
\end{eqnarray}

We know that the Cauchy problem (no boundaries)
for (\ref{Eq:dshifted1gen}-\ref{Eq:dshifted2gen}) is stable for any
value of the $\beta$ in both second and fourth-order accuracy \cite{C1}.

For the quarter space problem, it is convenient to introduce ghost
points, that is, we assume that the interior equations hold for all $j \ge
0$ and provide numerical prescriptions for the grid functions at the
grid points $j=-p/2, \ldots, -1$.  Stable discrete boundary conditions
are given in the following subsections.  The proofs follow in Section
\ref{Sec:Laplace}.

For completeness, and for comparison with the second-order wave
equation, we give second and fourth-order accurate boundary prescriptions
for the advection equation in \ref{Sec:advection}.

\subsection{Second order accuracy}

\subsubsection{Outflow boundary}

We start with the outflow case $\beta > 1$. The continuum problem does
not require any boundary condition, but at the discrete level a special
prescription is needed.
Third order extrapolation for $\phi_j$ and second order extrapolation
for $\Pi_j$, namely
\begin{eqnarray}
h^3 D_+^3 \phi_{-1} &=& 0, \\
h^2 D_+^2 \Pi_{-1} &=& 0,
\end{eqnarray}
give strong stability and second order convergence.  Interestingly,
the minimum order of extrapolation required for second order
convergence is not the same for the two grid functions $\phi_j$ and
$\Pi_j$.  The reason for this becomes clear in the convergence
analysis of subsection \ref{Sec:outflow}.  This is to be contrasted
with the result for fully first order symmetrizable hyperbolic
systems, in which second order extrapolation (or, equivalently, first
order one-sided differencing) for the outgoing characteristic
variables yields second order convergence.

\subsubsection{Time-like boundary}

When the shift satisfies $|\beta|<1$, one of the two characteristic
variables is entering the domain through the boundary.  We seek
discrete boundary conditions approximating
\begin{equation}
\Pi(0,t) - \p_x \phi(0,t) = g(t), \label{Eq:Sommerfeld2}
\end{equation}
which lead to strong stability and preserve the internal accuracy.
This is achieved by populating the ghost point $j=-1$ using
\begin{eqnarray}
&&\Pi_0 - D_0\phi_0 = g,\\
&&h^2D_+^2 \Pi_{-1} = 0,
\end{eqnarray}
or, explicitly, 
\begin{eqnarray}
\phi_{-1} &=& \phi_1 + 2h(g-\Pi_0),\\
\Pi_{-1} &=& 2\Pi_0 -\Pi_1.
\end{eqnarray}

\subsection{Fourth order accuracy}

\subsubsection{Outflow boundary}

In the outflow case, the extrapolation conditions
\begin{eqnarray}
h^5 D_+^5 \phi_{-1} &=& 0, \\
h^5 D_+^5 \phi_{-2} &=& 0, \\
h^4 D_+^4 \Pi_{-1} &=& 0, \\
h^4 D_+^4 \Pi_{-2} &=& 0,
\end{eqnarray}
lead to strong stability and fourth order convergence.

\subsubsection{Time-like boundary}

The prescriptions
\begin{eqnarray}
\Pi_0 - D^{(1)} \phi_0 &=& g, \label{Eq:shift4TL1}\\
h^5 D_+^5\phi_{-2} &=& 0,\label{Eq:shift4TL2}\\
h^4D_+^4\Pi_{-1} &=& 0,\label{Eq:shift4TL3}\\
h^4D_+^4\Pi_{-2} &=& 0,\label{Eq:shift4TL4}
\end{eqnarray}
where $D^{(1)}$ is defined in (\ref{Eq:D1}), give strong stability and
fourth order convergence.  (Increasing the extrapolation order by one
in both $\phi$ and $\Pi$ also gives fourth order convergence.)
Explicitly solving (\ref{Eq:shift4TL1}) and (\ref{Eq:shift4TL2}) for
$\phi_{-1}$ and $\phi_{-2}$ gives
\begin{eqnarray}
\phi_{-1} &=& 4(g-\Pi_0)h -\frac{10}{3} \phi_0 + 6 \phi_1 - 2\phi_2 +
\frac{1}{3} \phi_3, \\
\phi_{-2} &=& 20(g-\Pi_0)h - \frac{80}{3} \phi_0 + 40 \phi_1 -
15\phi_2 + \frac{8}{3} \phi_3.
\end{eqnarray}

\section{Proofs of strong stability and convergence}
\label{Sec:Laplace}

In this Section we use the Laplace transform method for difference
approximations as described in Chapter 12 of \cite{GKO} to prove
strong stability for second and fourth-order accurate discretisations
of the initial-boundary value problem for the shifted wave equation.
In order to apply the theorems of that reference, which assume that
hyperbolic problems are written in fully first order form (see
equation (12.1.11) of \cite{GKO}), we need to
perform a {\em discrete reduction to first order} \cite{CHH}.

\subsection{Second order accuracy}

\subsubsection{Outflow boundary}
\label{Sec:outflow}

We consider the semi-discrete quarter space problem for the outflow
case ($\beta >1$)
\begin{eqnarray}
&&\frac{\rmd}{\rmd t}\phi_j = \beta D_0\phi_j + \Pi_j + F^{\phi}_{j}\label{Eq:sd1}, \\
&&\frac{\rmd}{\rmd t}\Pi_j = \beta D_0 \Pi_j + D_+D_-\phi_j + F^{\Pi}_{j}\label{Eq:sd2}, \\
&&\phi_j(0) = f^{\phi}_j\label{Eq:sd3}, \\
&&\Pi_j(0) = f^{\Pi}_j\label{Eq:sd4}, \\
&&h^{q_2+1} D_+^{q_2+1} \phi_{-1} = g^{\phi}\label{Eq:extrap_phi}, \\
&&h^{q_1} D_+^{q_1} \Pi_{-1} = g^{\Pi}\label{Eq:extrap_Pi}, \\
&&\|\Pi\|^2_h +\|D_+\phi\|_h^2 < \infty\label{Eq:sd7},
\end{eqnarray}
where $j=0,1,2,\ldots$, $\| u \|_h^2 = \sum_{j=0}^{\infty} |u_j|^2
h$, and $q_1$ and $q_2$ are non-negative integers. In
practice one would choose $g^{\phi}=g^{\Pi}=0$, but in the analysis
that follows we will need the inhomogeneous case.  A
first order reduction of the problem obtained by introducing the
grid function $X_j = D_+\phi_j$ gives
\begin{eqnarray}
&&\frac{\rmd}{\rmd t}\Pi_j = \beta D_0 \Pi_j + D_-X_j + F^{\Pi}_{j}\label{Eq:red1}, \\
&&\frac{\rmd}{\rmd t}X_j = \beta D_0X_j + D_+\Pi_j + F^{X}_{j}\label{Eq:red2}, \\
&&\Pi_j(0) = f^{\Pi}_j\label{Eq:red3}, \\
&&X_j(0) = f^X_j\label{Eq:red4}, \\
&&h^{q_1} D_+^{q_1} \Pi_{-1} = g^{\Pi}\label{Eq:red5}, \\
&&h^{q_2} D_+^{q_2} X_{-1} = g^{X}\label{Eq:red6},
\end{eqnarray}
where $j=0,1,2,\ldots$, $F^X_j = D_+F^{\phi}_j$, $f^X_j =
D_+f^{\phi}_j$, and $g^{X} = g^{\phi}/h$. The auxiliary
constraint $C_j \equiv X_j - D_+\phi_j$ satisfies the homogeneous
system of ODEs
\begin{eqnarray}
&&\frac{\rmd}{\rmd t} C_j = \beta D_0 C_j,\qquad j = 0, 1, 2, \ldots,\\
&& C_j (0) = 0,\\
&& h^{q_2} D_+^{q_2} C_{-1} = 0,
\end{eqnarray}
and therefore vanishes identically.

We want to show that the semi-discrete initial-boundary value problem
(\ref{Eq:red1}-\ref{Eq:red6}) is strongly stable and second order
convergent if $q_1,q_2\ge 2$. Since the evolution equation
for $\phi_j$ only involves lower order terms ($D_0\phi_j$ can be
expressed as a combination of $X_j$ and $X_{j-1}$), it can be ignored
in the analysis that follows.

The proof of stability is divided into three steps. We first estimate
the solution of the problem with $F=0$ and $f=0$ in terms of the
boundary data by checking that the Kreiss condition, inequality
(\ref{Eq:Kreisscond}), is satisfied.  We then estimate the solution of
an auxiliary problem with modified homogeneous boundary conditions.
Finally, we combine the two estimates to obtain the strong stability
estimate, inequality (\ref{Eq:strongstab}), for the original problem.
In its final step the proof relies on plotting a function of a
complex variable to show that it is bounded.

\paragraph*{Step 1. By estimating the solution of the $F=0$, $f=0$ problem
  near the boundary using the Kreiss condition {\em(\ref{Eq:Kreisscond})} we obtain the estimate {\em (\ref{Eq:estimateg})}.}

Let $F=0$ and $f=0$ in Eqs.~(\ref{Eq:red1}-\ref{Eq:red6}). We want
to show that
\begin{equation}
\| \Pi(t)\|_h^2 + \|X(t)\|_h^2 \le {\rm const} \int_0^t (|
g^{\Pi}(\tau) |^2  + |g^X(\tau)|^2 ) \, \rmd\tau \label{Eq:estimateg}.
\end{equation}
Observing that 
\begin{eqnarray}
\fl\frac{\rmd}{\rmd t} (\| \Pi(t)\|_h^2 + \|X(t)\|_h^2) \label{Eq:estimateu} = -\beta (\Pi_0\Pi_{-1}
+ X_0 X_{-1}) - 2 \Pi_0 X_{-1} \\
\le {\rm const} \sum_{j=-1}^0 (|\Pi_j|^2 + |X_j|^2), \nonumber 
\end{eqnarray}
it is clear that if we can estimate the solution near the boundary
(i.e., at $j=-1,0$) in terms of the boundary data, we recover the
estimate (\ref{Eq:estimateg}).  We show that this is possible by
explicitly solving the Laplace transformed problem
\begin{eqnarray}
\tilde{s} \hat{\Pi}_j &=& \beta (\hat{\Pi}_{j+1} - \hat\Pi_{j-1})/2 +
\hat X_j - \hat X_{j-1}\label{Eq:Lap1}, \\
\tilde{s} \hat X_j &=& \beta (\hat{X}_{j+1} - \hat X_{j-1})/2 +
\hat \Pi_{j+1} - \hat \Pi_{j} \label{Eq:Lap2}, \\
&&h^{q_1} D_+^{q_1} \hat \Pi_{-1} = \hat g^{\Pi}\label{Eq:Lap3}, \\
&&h^{q_2} D_+^{q_2} \hat X_{-1} = \hat g^{X}\label{Eq:Lap4}, \\
&&\| \hat{\Pi}\|_h^2 + \| \hat X\|_h^2 < \infty\label{Eq:Lap5},
\end{eqnarray}
where $\hat u(s) = \int_0^{+\infty} \rme^{-st} u(t) \rmd t$ and $\tilde s =
sh$.

Eqs.~(\ref{Eq:Lap1}) and (\ref{Eq:Lap2}) form a system of
difference equations. The characteristic equation
associated with it, obtained by looking for solutions of the
form $\hat \Pi_j = k^j \hat \Pi_0$, $\hat X_j = k^j \hat X_0$, is
\begin{eqnarray}
&&\left( \tilde s - \frac{\beta}{2}(k-k^{-1})\right)^2 -
\frac{(k-1)^2}{k} = 0, \label{Eq:charequation}
\end{eqnarray}
a polynomial of degree 4 in $k$.  For $\Re(\tilde{s})>0$ there are no
solutions with $|k|=1$.  If $k=\rme^{\rmi\xi}$ with $\xi \in \mathbb{R}$
was a solution we would have $\Re(\tilde s) = 0$, which is a
contradiction.  Observing that the roots are continuous functions of
$\tilde s$ and for large values of $|\tilde s|$ we have $\tilde s \simeq
\beta/(2k) \pm 1/\sqrt{k}$, we conclude that for $\Re(\tilde s)> 0$
there are two and only two solutions, $k_1$ and $k_2$, inside the unit
circle.  For $\tilde s = 0$ the four roots are
\begin{equation}
k_{1,2} = \frac{2-\beta^2\pm 2\rmi
  \sqrt{\beta^2-1}}{\beta^2}, \quad k_{3,4} = 1.
\end{equation}
Since, for small $|\tilde s|$, $k_{3,4} = 1 + (\beta\pm 1)^{-1} \tilde
s + \Or(|\tilde s|^2)$, the roots $k_1$ and $k_2$ are those which are
inside the unit circle for $\Re (\tilde s)>0$.

The general solution of the difference equation
(\ref{Eq:Lap1}-\ref{Eq:Lap2}) satisfying $\|\hat{\Pi}\|^2_h + \|
\hat{X}\|^2_h < \infty$ can be written as
\begin{eqnarray}
\hat{\Pi}_j &=& \sigma_1 \tilde{s}_1 k_1^j +
\sigma_2 \tilde{s}_2 k_2^j \label{Eq:Pihat}, \\
\hat{X}_j &=& \sigma_1 (k_1-1) k_1^j + \sigma_2(k_2-1) k_2^j \label{Eq:Xhat},
\end{eqnarray}
where $\tilde{s}_{1,2} = \tilde{s} -\frac{\beta}{2} (k_{1,2}
-k_{1,2}^{-1})$.  Note that since we will be constructing the explicit
solution, the case $k_1 = k_2$ does not require special treatment.
Inserting (\ref{Eq:Pihat}-\ref{Eq:Xhat}) into the boundary
conditions (\ref{Eq:Lap3}-\ref{Eq:Lap4}) gives rise to a $2\times
2$ system in $\sigma_1$ and $\sigma_2$
\begin{eqnarray}
&&\sigma_1 \tilde s_1 (k_1-1)^{q_1}k_1^{-1} + \sigma_2
  \tilde s_2 (k_2-1)^{q_1}k_2^{-1} = \hat g^{\Pi}, \\  
&&\sigma_1 (k_1-1)^{q_2+1}k_1^{-1} +
\sigma_2 (k_2-1)^{q_2+1}k_2^{-1} = \hat g^X .
\end{eqnarray}
If the coefficient matrix $C(\tilde s)$ multiplying
$(\sigma_1,\sigma_2)^{\rm T}$ is non-singular for $\Re(\tilde{s})>0$, we
can solve for $\sigma_1$ and $\sigma_2$ and substitute into
(\ref{Eq:Pihat}) and (\ref{Eq:Xhat}), and obtain an explicit solution
of the Laplace transformed problem, which we write in the form
\begin{eqnarray}
\hat \Pi_j &=& \sum_{k=\Pi,X} c_{jk}^{\Pi} \hat g^k,\label{Eq:gensol1} \\
\hat X_j &=& \sum_{k=\Pi,X} c_{jk}^X \hat g^k. \label{Eq:gensol2}
\end{eqnarray}

\begin{figure}[ht]
\begin{center}
\includegraphics*[width=7cm]{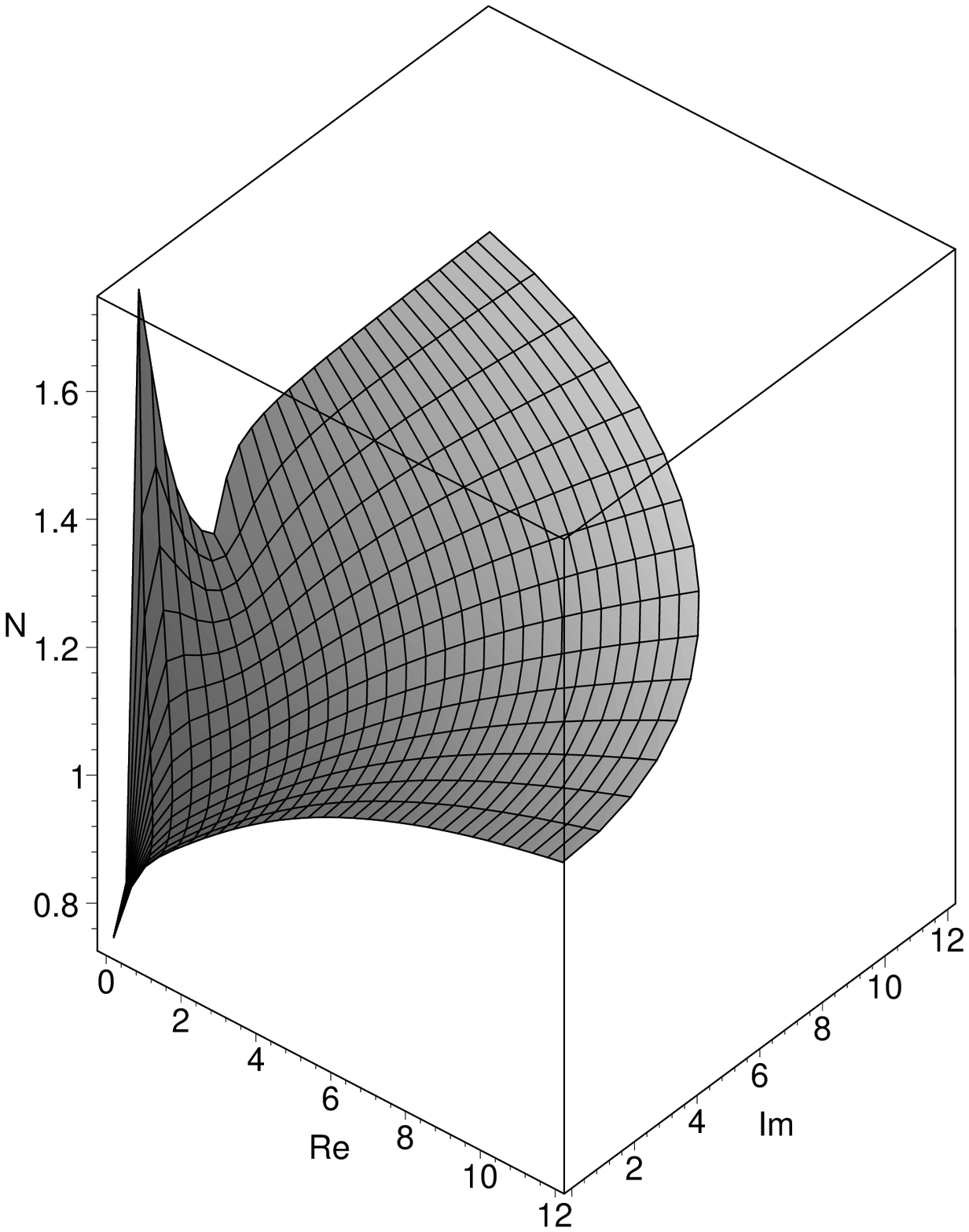}
\includegraphics*[width=7cm]{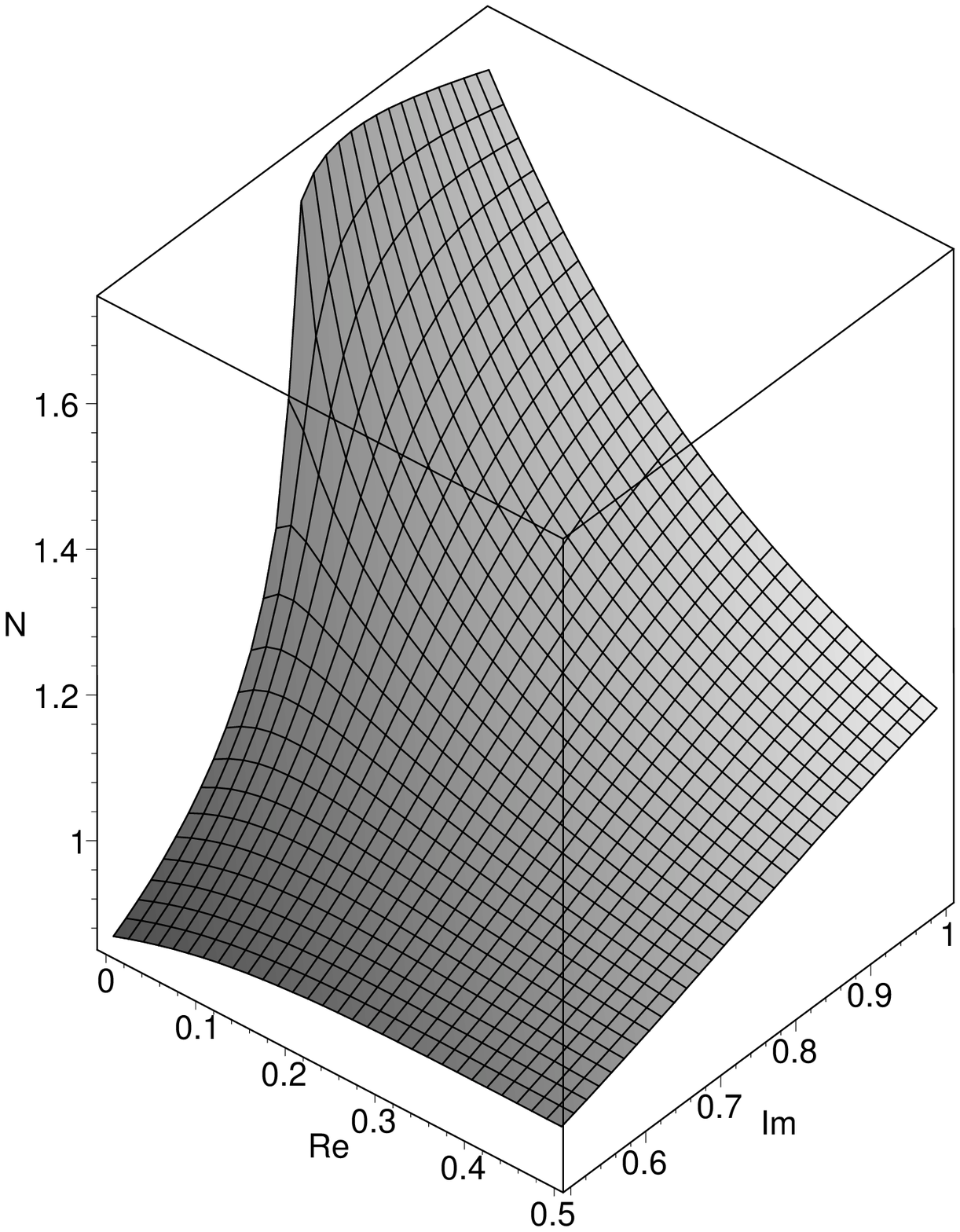}
\caption{On the left we plot $N(\tilde s)$ for $\tilde s \in S
  \cap \{{\rm Im}(\tilde s) \ge 0\}$.  Since this function is
  symmetric across the real axis, we only need to display the region
  with non-negative imaginary part.  The Kreiss condition is satisfied
  for the initial-boundary value problem (\ref{Eq:sd1}-\ref{Eq:sd7})
  with $\beta = 2$ and $q_1=q_2=2$.  The spike that appears at a point
  near ${\rm Re}(\tilde s) = 0$ is bounded.  This is illustrated on
  the right, where we have increased the resolution of the plot in a
  neighbourhood of this point. The function $N(\tilde s)$ is
  continuous but not differentiable on ${\rm Re} (\tilde s) =0$ due to
  branch cuts in its definition.}
\label{Fig:shifted_beta2Kr_2nd}
\end{center}
\end{figure}

To verify the Kreiss condition, namely that 
\begin{equation}
\sum_{j=-1}^0 \left(| \hat \Pi_j |^2 + | \hat X_j |^2\right) \le K ( |
\hat g^{\Pi}|^2 + | \hat g^X|^2), \label{Eq:Kreisscond}
\end{equation}
we numerically compute the coefficients
$c_{jk}^{\Pi}$, $c_{jk}^X$, for $j=-1,0$ and $k=X,\Pi$, and plot the
quantity
\begin{equation}
N = \left(\sum_{\tiny\begin{array}{c}
j=-1,0, \\
k=\Pi, X
\end{array}} ( |c^{\Pi}_{jk}|^2 + |c^{X}_{jk}|^2)\right)^{1/2}. \label{Eq:N}
\end{equation}
We restrict our attention to the $q_1=q_2=2$ case (same order of
extrapolation for $\Pi_j$ and $X_j$).  Since for $|\tilde{s}| \ge C_0$
we have that $|\hat\Pi_j|^2 + |\hat X_j|^2 \le K_0( | \hat g^{\Pi}|^2
+ | \hat g^X|^2)$, see Lemma 12.2.2 of \cite{GKO}, we only need to
consider the compact set $S = \{ \tilde s \in {\mathbb C}: |\tilde s|
\le C_0,\,\Re(\tilde s) \ge 0\}$. From the chain of inequalities
\begin{eqnarray*}
\fl &&\| \hat\Pi \|^2_h + \| \hat X \|_h^2 = \frac{1}{|\tilde s|^2} \left( \|
h\beta D_0 \hat\Pi + h D_-\hat X \|_h^2 + \| h \beta D_0 \hat X + h D_+ \hat\Pi\|^2_h\right)\\
\fl && \le \frac{2}{|\tilde s|^2} \left( \beta^2 \| h D_0 \hat\Pi\|_h^2 + \| h
D_-\hat X \|_h^2 + \beta^2 \| h D_0 \hat X \|_h^2 + \| h D_+ \hat
\Pi\|^2_h\right)\\
\fl && \le  \frac{2}{|\tilde s|^2} \left[ \beta^2 \left( \| \hat\Pi \|_h^2 +
  \frac{1}{2} | \hat\Pi_{-1} |^2 h\right) + 4 \| \hat X \|_h^2 + 2 | \hat X_{-1} |^2
  h + \beta^2 \left( \| \hat X \|_h^2 + \frac{1}{2} | \hat X_{-1} |^2 h\right) +
  4\| \hat\Pi \|_h^2 \right]\\
\fl && \le \frac{14}{| \tilde s|^2} (4+\beta^2) \left( \| \hat\Pi \|_h^2 +
\| \hat X \|_h^2 + | g^{\Pi}|^2 h + | g^{X}|^2 h\right),
\end{eqnarray*}
where in the last inequality we used the boundary conditions
(\ref{Eq:Lap3}-\ref{Eq:Lap4}), we see that for $\beta = 2$ we can take
$C_0 = 12$.  In \fref{Fig:shifted_beta2Kr_2nd} we plot $N(\tilde s)$
where $\tilde s \in S$.

Using Parseval's relation and the fact that the
solution at time $t_1$ does not depend on the boundary data at time
$t_2>t_1$, the Kreiss condition implies the estimate in physical space
\begin{equation}
\int_0^t \sum_{j=-1}^0(|\Pi_j|^2 + |X_j|^2 ) \,\rmd\tau \le K \int_0^t (
| g^{\Pi}|^2 + | g^X|^2) \,\rmd\tau.
\end{equation}
Combining this with the integral of inequality (\ref{Eq:estimateu})
gives the desired estimate (\ref{Eq:estimateg}).

\paragraph*{Step 2: We estimate the solution of a problem with modified
homogeneous boundary conditions in terms of the initial data and
forcing term, inequalities {\em (\ref{Eq:normwestimate})} and {\em
(\ref{Eq:wjestimate})}.}

So far we have shown that, for vanishing initial data and in the
absence of a forcing term, the solution can be estimated in terms of
the boundary data.  We now consider the auxiliary problem
\begin{eqnarray}
&&\frac{\rmd}{\rmd t} \Pi_j  = \beta D_0 \Pi_j + D_- X_j + F^{\Pi}_j,\\
&&\frac{\rmd}{\rmd t} X_j =  \beta D_0 X_j + D_+ \Pi_j + F^X_j,\\
&&\Pi_j(0) = f^{\Pi}_{j}, \quad X_j(0) = f^X_{j},\\
&&\Pi_{-1} = \frac{2}{\beta}\left(\Pi_0 - \frac{2}{\beta} X_0\right),\\
&& X_{-1} = \frac{2}{\beta} X_0,\\
&&\| \Pi \|_h^2 + \| X \|_h^2 < \infty,
\end{eqnarray}
where the boundary conditions were chosen so that a direct estimate in
physical space can be obtained.  The estimate
\begin{eqnarray}
\fl\frac{\rmd}{\rmd t} (\|\Pi\|_h^2 +\|X\|_h^2) \le - 2(|\Pi_0|^2 +|X_0|^2)+\|\Pi\|_h^2 +\|X\|_h^2 + \|F^{\Pi}\|_h^2 +\|F^X\|_h^2
\end{eqnarray}
implies 
\begin{equation}
\| \Pi (t) \|_h^2 + \| X(t) \|_h^2 \le {\rm const} \left( \| f \|_h^2 +
\int_0^t \| F(\tau) \|_h^2 \,\rmd\tau\right) \label{Eq:normwestimate}
\end{equation}
and
\begin{eqnarray}
\fl\int_0^t (|\Pi_j|^2 + |X_j|^2 ) \,\rmd\tau \le {\rm const} \left( \| f^{\Pi}\|_h^2 + \|
f^X \|_h^2 + \int_0^t (\|F^{\Pi}\|_h^2 +\|F^X\|_h^2)\rmd \tau\right)
\end{eqnarray}
for $j=-1,0$.  Since our interior scheme uses the same number of
grid points in each side, one can show (Lemma 12.2.10 of \cite{GKO})
that for every fixed $j$
\begin{equation}
\int_0^{\infty} ( |\Pi_j|^2 + |X_j|^2 ) \,\rmd t \le {\rm const} \left(
\|f\|_h^2 + \int_0^{\infty} \| F(t) \|_h^2 \,\rmd t\right) \label{Eq:wjestimate},
\end{equation}
where $f = (f^{\Pi},f^{X})^{\rm T}$ and $F=(F^{\Pi},F^X)^{\rm T}$.

\paragraph*{Step 3: Using the estimates of Steps 1 and 2 we derive the
estimate {\em (\ref{Eq:strongstab})} showing strong stability.}

If we denote by $\Pi^{\rm a}_j$ and $X^{\rm a}_j$ the solution of the
auxiliary problem of Step 2, and by $\Pi_j$ and $X_j$ the solution of
the original problem, we see that the differences $\bar \Pi_j = \Pi_j
- \Pi_j^{\rm a}$ and $\bar X_j = X_j - X_j^{\rm a}$ satisfy
\begin{eqnarray}
&&\frac{\rmd}{\rmd t} \bar \Pi_j = \beta D_0 \bar \Pi_j + D_- \bar X_j
, \\
&&\frac{\rmd}{\rmd t} \bar X_j = \beta D_0 \bar X_j + D_+ \bar \Pi_j, \\
&&\bar \Pi_j (0) = 0 , \\
&&\bar X_j (0) = 0 , \\
&&h^{q_1} D_+^{q_1} \bar \Pi_{-1} = g^{\Pi} - h^{q_1} D_+^{q_1}
\bar \Pi_{-1}^{\rm a}, \\
&&h^{q_2} D_+^{q_2} \bar X_{-1} = g^{X} - h^{q_2} D_+^{q_2}
\bar X_{-1}^{\rm a}.
\end{eqnarray}
Using (\ref{Eq:estimateg}), (\ref{Eq:wjestimate}) and
(\ref{Eq:normwestimate}) we obtain the estimate
\begin{eqnarray}
\fl \| \Pi(t)\|_h^2 + \| X(t) \|_h^2 \le \label{Eq:strongstab} 2 (\| \bar\Pi(t)\|_h^2 + \| \bar X(t) \|_h^2 +\|
  \Pi^{\rm a}(t)\|_h^2 + \| X^{\rm a}(t) \|_h^2 )\le\\ 
{\rm const} \left( \| f\|_h^2  + \int_0^t \left( \|F(\tau)\|_h^2 +
|g(\tau)|^2\right)\,\rmd\tau\right)\nonumber,
\end{eqnarray}
where $g= (g^{\Pi},g^X)^{\rm T}$.  This inequality proves strong
stability.  Reintroducing the evolution equation for $\phi_j$, we have
an estimate with respect to the $D_+$-norm
\begin{eqnarray}
\fl\|\phi(t) \|_h^2 + \| \Pi(t)\|_h^2 + \| D_+\phi(t)\|_h^2 \le\label{Eq:strongstabestimate} {\rm const} \Big( \|f^{\phi} \|_h^2 + \| f^{\Pi}\|_h^2 + \|
  D_+f^{\phi}\|_h^2 + \\
\int_0^t \big( \| F^\phi(\tau)\|_h^2 +\| F^{\Pi}(\tau)\|_h^2 + \|
D_+F^{\phi}(\tau)\|_h^2 + |g^{\Pi}(\tau)|^2 + |g^{\phi}(\tau)/h|^2 \big)\rmd\tau\Big).\nonumber
\end{eqnarray}

\paragraph*{Proof of convergence:} Having shown strong stability it is
straightforward to prove convergence. We only need an estimate for the
error.  Defining the errors $e^\phi_j(t) = \phi_j(t) -\phi(x_j,t)$,
$e^\Pi_j(t) = \Pi_j(t) - \Pi(x_j,t)$ we see that they satisfy the
initial-boundary value problem
\begin{eqnarray}
&&\frac{\rmd}{\rmd t} e^\phi_j = \beta D_0 e^\phi_j + e^\Pi_j + F_j^\phi\label{Eq:err1}, \\
&&\frac{\rmd}{\rmd t} e^\Pi_j = \beta D_0 e^\Pi_j + D_+D_- e^{\phi}_j +
F^\Pi_j \label{Eq:err2}, \\
&&e^\phi_j(0) = 0\label{Eq:err3}, \\
&&e^\Pi_j(0) = 0\label{Eq:err4}, \\
&&h^{q_2+1} D_+^{q_2+1} e^\phi_{-1}  = - h^{q_2+1}  \phi^{(q_2+1)}(0,t)+\Or(h^{q_2+2})\label{Eq:err5}, \\
&&h^{q_1} D_+^{q_1} e^\Pi_{-1} = -h^{q_1} \Pi^{(q_1)}(0,t)+\Or(h^{q_1+1})\label{Eq:err6},
\end{eqnarray}
where $F_j^\phi = \Or(h^2)$ and $F_j^\Pi= \Or(h^2)$.  We perform a
discrete reduction by introducing the quantity $e^X_j \equiv D_+ e^\phi_j$ which satisfies
\begin{eqnarray}
&&\frac{\rmd}{\rmd t} e^X_j = \beta D_0 e^X_j + D_+ e^{\Pi}_j + F^X_j,\label{Eq:err7} \\
&&e^X_j(0) = 0,\label{Eq:err8} \\
&&h^{q_2} D_+^{q_2} e^X_{-1} = - h^{q_2} \phi^{(q_2+1)}(0,t)+\Or(h^{q_2+1}).\label{Eq:err9}
\end{eqnarray}
Note that since $F^{\phi}_j = \beta(D_0\phi(x_j,t) - \phi_x(x_j,t))$,
we have $F_j^X \equiv D_+ F^\phi_j = \Or(h^2)$.

The strong stability estimate (\ref{Eq:strongstabestimate}) applied to
(\ref{Eq:err2}), (\ref{Eq:err4}), (\ref{Eq:err6}-\ref{Eq:err9}),
guarantees that the scheme is second-order convergent, i.e.
\begin{equation}
 (\|e^{\phi}\|_h^2 + \| e^{\Pi}\|_h^2 + \|D_+e^{\phi}\|_h^2)^{1/2} \le
 \Or(h^2), \label{Eq:errorestimate}
\end{equation}
provided that $q_1,q_2 \ge 2$.  In this analysis we have
implicitly assumed that the initial data and forcing terms are exact.
However, this assumption can be easily replaced by the requirement
that the initial data errors for $e^{\Pi}_j$ and $e^X_j$ are of order
$h^2$ (note that this means that $D_+e^{\phi}_j = \Or(h^2)$) and that
the errors in the forcing terms $F^{\Pi}_j$ and $F^X_j$ are of order
$h^2$.  An immediate consequence of (\ref{Eq:errorestimate}) is that
we have convergence with respect to the discrete $L_2$-norm,
$(\|\phi\|_h^2 + \|\Pi\|_h^2)^{1/2}$.

We have also studied the case $q_1 = q_2 + 1 = 3$ (same order of
extrapolation for $\phi_j$ and $\Pi_j$).  The Kreiss condition can be
verified directly and the rest of the stability proof applies.

\subsubsection{Time-like boundary}

In this Subsection we prove stability for the shifted wave equation
problem with a discretization of the Sommerfeld boundary condition,
equation \eref{Eq:Sommerfeld2}.  The case with zero shift was discussed in
Appendix A of \cite{C2} using the discrete energy method.  Here we
focus on the $0\neq |\beta|<1$ case.

We consider the same semi-discrete evolution system
(\ref{Eq:sd1}-\ref{Eq:sd4}) and (\ref{Eq:sd7}) with boundary
conditions
\begin{eqnarray}
\Pi_0 - D_0\phi_0 = g^X,\label{Eq:bcshifted1} \\
h^qD_+^q \Pi_{-1} = g^{\Pi}.\label{Eq:bcshifted2}
\end{eqnarray}
In applications one would set $g^X$ equal to the boundary data $g(t)$ of the
continuum problem and $g^{\Pi}=0$.  The discrete reduction of
(\ref{Eq:bcshifted1}) is 
\begin{equation}
\Pi_0 - \frac{1}{2} (X_0 +X_{-1}) = g^X, \label{Eq:bcshifted1X}
\end{equation}
which implies the boundary condition $C_{-1} = -C_0$ for the auxiliary
constraint. 

The proof of strong stability proceeds as in the outflow case.  We
only need to show that the Kreiss condition is satisfied.  Inserting
$\Pi_j(t) = \rme^{st} k^j \hat\Pi_0$ and $X_j(t) = \rme^{st} k^j \hat X_0$
into the scheme, and looking for non trivial solutions gives the
characteristic equation (\ref{Eq:charequation}).  For small $|\tilde
s|$ the roots are
\begin{eqnarray}
k_1 &=& \frac{2-\beta^2-2\sqrt{1-\beta^2}}{\beta^2} + \Or(|\tilde s|), \\
k_2 &=& 1 + \frac{1}{\beta-1}\tilde s + \Or(|\tilde s|^2), \\
k_3 &=& \frac{2-\beta^2+2\sqrt{1-\beta^2}}{\beta^2} + \Or(|\tilde s|), \\
k_4 &=& 1 + \frac{1}{\beta+1}\tilde s + \Or(|\tilde s|^2).
\end{eqnarray}
For $\Re(\tilde s)>0$ the roots $k_1$ and $k_2$ have magnitude smaller
than 1 and the remaining two have magnitude greater than 1.  The
requirement $\|\hat \Pi\|^2_h + \|\hat X\|_h^2 < \infty$ implies that
the general solution will have the form (\ref{Eq:Pihat}) and
(\ref{Eq:Xhat}), where the parameters $\sigma_1$ and $\sigma_2$ can be
determined by imposing the boundary conditions
\begin{eqnarray}
\fl\sigma_1 \left( s-\frac{1}{2} (1+\beta) (k_1-k_1^{-1})\right) -
\frac{\sigma_2}{2} (1+\beta) \left( 1+ \frac{1}{k_1k_2}\right) = \hat g^X,\\
\fl\sigma_1 \tilde s_1 (k_1-1)^{q_1}k_1^{-1} + \frac{\sigma_2}{k_2-k_1}
  \left(\tilde s_2 (k_2-1)^{q_1}k_2^{-1} - \tilde s_1
  (k_1-1)^{q_1}k_1^{-1}\right) = \hat g^{\Pi}.
\end{eqnarray}

The solution, if it exists, has the form
(\ref{Eq:gensol1}-\ref{Eq:gensol2}).  Again, we verify the Kreiss
condition by plotting the quantity (\ref{Eq:N}) for $\tilde s \in S$
and verifying that is is bounded.  

Having established that the estimate (\ref{Eq:estimateg}) holds, we
use the same auxiliary problem used in the outflow case, Step 2,
giving the estimates (\ref{Eq:normwestimate}) and
(\ref{Eq:wjestimate}).  Hence, we have strong stability.  Convergence
follows by observing that the error equation associated with
(\ref{Eq:bcshifted1}) is
\begin{equation}
e^{\Pi}_0 -D_0 e^{\phi}_0 = \frac{h^2}{6} \phi'''(0,t) + \Or(h^4).
\end{equation}

\subsection{Fourth order accuracy}

\subsubsection{Outflow boundary}

The fourth order accurate standard discretization of
the shifted wave equation is given by 
\begin{eqnarray}
&&\frac{\rmd}{\rmd t}\phi_j = \beta D^{(1)}\phi_j + \Pi_j + F^{\phi}_{j},\label{Eq:sd41} \\
&&\frac{\rmd}{\rmd t}\Pi_j = \beta D^{(1)} \Pi_j + D^{(2)}\phi_j +
  F^{\Pi}_{j},\label{Eq:sd42}
\end{eqnarray}
where $j=0,1,2,\ldots$ and
\begin{eqnarray}
\fl D^{(1)}u_j \equiv D_0\left(  1 - \frac{h^2}{6}D_+D_-\right)u_j= (-u_{j+2}+8u_{j+1}-8u_{j-1} + u_{j-2})/(12h),\\
\fl D^{(2)}u_j \equiv D_+D_- \left( 1 - \frac{h^2}{12}D_+D_-\right)u_j
\\
= (-u_{j+2}+16u_{j+1} - 30u_j +16u_{j-1} - u_{j-2})/(12h^2). \nonumber
\end{eqnarray}
We consider the outflow case first ($\beta > 1$), with the 
boundary conditions
\begin{eqnarray}
&&h^5 D_+^5 \phi_{-1} = g^{\phi}_1, \quad h^5 D_+^5 \phi_{-2} = g^{\phi}_2,\label{Eq:extrap41} \\
&&h^4 D_+^4 \Pi_{-1} = g^{\Pi}_1, \quad h^4 D_+^4 \Pi_{-2} =
  g^{\Pi}_2.\label{Eq:extrap42}
\end{eqnarray}

As in the second order accurate case, we need to perform a
discrete reduction to first order.  For this purpose it is convenient
to introduce a grid function $\tilde X_j$, which is a suitable linear
combination of $D_+\phi_j$ and $D_-\phi_j$.  The choice of the
particular combination is determined by the following two
observations.  First, the operator $D^{(2)}$ can be decomposed as
\begin{equation}
D^{(2)} = \tilde D_+ \tilde D_-,
\end{equation}
where $\tilde D_{\pm} = (1 \mp \alpha h D_{\mp}) D_{\pm} =
(1-\alpha)D_{\pm} + \alpha D_{\mp}$ and $\alpha$
is a root of the quadratic equation $\alpha^2 - \alpha - 1/12 =
0$. Second, in the absence of boundaries, the following discrete
energy,
\begin{equation}
\| \Pi \|_h^2 + \| D_+\phi \|_h^2 + \frac{h^2}{12} \| D_+D_-\phi\|_h^2,
\end{equation}
which is equivalent to $\| \Pi \|_h^2 + \| D_+\phi\|_h^2$, is
conserved and it can be written as\footnote{To show that 
$\|D_+ \phi_j
- \alpha h D_+D_-\phi_j\|_h^2 = \| D_+ \phi\|_h^2 + h^2/12\, \|
D_+D_-\phi\|_h^2$ one can use the identity $hD_+D_- = D_+ - D_-$ and
$\alpha^2 - \alpha - 1/12 = 0$.}
\begin{equation}
\| \Pi \|_h^2 + \| \tilde X \|_h^2,
\end{equation}
where $\tilde X_j = \tilde D_+ \phi_j$.  

These two facts suggest
introducing the discrete variable $\tilde X_j$, leading to the
(interior) discrete reduction
\begin{eqnarray}
\frac{\rmd}{\rmd t}\Pi_j &=& \beta D^{(1)} \Pi_j + \tilde D_-\tilde X_j + F^{\Pi}_{j}\label{Eq:red41}, \\
\frac{\rmd}{\rmd t}\tilde X_j &=& \beta D^{(1)}\tilde X_j + \tilde D_+\Pi_j + F^{X}_{j}\label{Eq:red42}, \\
\Pi_j(0) &=& f^{\Pi}_j\label{Eq:red43}, \\
X_j(0) &=& f^X_j\label{Eq:red44},
\end{eqnarray}
where $F^X_j = \tilde D_+F^{\phi}_j$, $f^X_j = \tilde D_+f^{\phi}_j$.

We need to translate the extrapolation boundary conditions for $\phi$
and $\Pi$ in terms of boundary conditions for the new variables, $\Pi$
and $\tilde X$.  To do this, we go back to the original system
(\ref{Eq:sd41}-\ref{Eq:sd42}) and eliminate the variables
$\phi_{-1}$, $\phi_{-2}$, $\Pi_{-1}$, and $\Pi_{-2}$ using the
boundary conditions (\ref{Eq:extrap41}-\ref{Eq:extrap42}).
Defining $\tilde X_j= \tilde D_+\phi_j$ for $j\ge 0$ as before, where
in $\tilde X_0$ the grid function $\phi_{-1}$ is eliminated using
(\ref{Eq:extrap41}) but with $g^{\phi}_1 = 0$, we can eliminate each
occurrence of $\phi_j$ in terms of $\tilde X_j$.  The evolution
equations for the $\tilde X_j$ variables can be computed by taking
appropriate combinations of the evolution equations of the $\phi_j$
variables.  This leads to a semi-discrete problem for $\Pi_j$ and
$\tilde X_j$, $j \ge 0$.  For $j\ge 2$ the equations have the form
(\ref{Eq:red41}-\ref{Eq:red42}), with the exception that
\begin{equation}
F^X_2 =
\tilde D_+ F^{\phi}_2 + \alpha\beta g^{\phi}_1/(12h^2). \label{Eq:FX2}
\end{equation}
However, for $j=0,1$ they are more complicated.  To analyze the
stability of the system we reintroduce ghost points.  By setting the
evolution equations near the boundary to be formally equal those at
the interior, for $j=0,1$, we obtain the prescriptions
\begin{eqnarray}
\fl \Pi_{-1} = 4 \Pi_{0}-6 \Pi_{1}+4
\Pi_{2}-\Pi_{3}+g^{\Pi}_1 + \frac{g^X_1}{\beta} ,\label{Eq:ghostPX1} \\ 
\fl \Pi_{-2} = 10 \Pi_{0}-20 \Pi_{1}+15 \Pi_{2}-4 \Pi_{3}+4
g^{\Pi}_1+g^{\Pi}_2 + \frac{2(7-6\alpha)}{\beta}g^X_1,\label{Eq:ghostPX2} \\ 
\fl \tilde X_{-1} = 4 \tilde X_{0}-6 \tilde X_{1}+4 \tilde X_{2}-\tilde
X_{3}+(1-5\alpha) g^X_1+\alpha g^X_2,\label{Eq:ghostPX3} \\
\fl \tilde X_{-2} = \frac{-137+132 \alpha}{12\alpha-13} \tilde
X_{0}-\frac{-289+288 \alpha}{12\alpha-13} \tilde X_{1}+\frac{-241+252
  \alpha}{12\alpha-13} \tilde X_{2}-2 \frac{-43+48
  \alpha}{12\alpha-13} \tilde X_{3}\label{Eq:ghostPX4}\\+\frac{12 \alpha-11}{12\alpha-13}
\tilde X_{4} +\frac{1}{12\alpha-13} \tilde X_{5}
+12\frac{\alpha-1}{\beta(12\alpha-13)} \Pi_{0}-48
\frac{\alpha-1}{\beta(12\alpha-13)} \Pi_{1}\nonumber \\ +72
\frac{\alpha-1}{\beta(12\alpha-13)} \Pi_{2}-48
\frac{\alpha-1}{\beta(12\alpha-13)} \Pi_{3}+12\frac{
  \alpha-1}{\beta(12\alpha-13)} \Pi_{4}\nonumber \\ 
+\frac{2(6-6\alpha+53\alpha\beta^2-55\beta^2)}{\beta^2(12\alpha-13)}
g^X_1+\frac{\alpha-2}{12\alpha-13} g^X_2-12 
\frac{\alpha-1}{\beta(12\alpha-13)}g^{\Pi}_1.\nonumber
\end{eqnarray}
where $g^X_k = g^{\phi}_k/h$.  The forcing terms near the boundary are
$F^X_1 = \tilde D_+ F^{\phi}_1$ and $F^X_0 = \tilde D_+ F^{\phi}_0$
(assuming $F^{\phi}_{-1}$ to be defined via $h^5 D_+^5 F^{\phi}_{-1} =
0$).

We now need to show that the semi-discrete initial-boundary value
problem (\ref{Eq:red41}-\ref{Eq:ghostPX4}) is strongly stable and
fourth order convergent.  Neglecting forcing terms, we have the
estimate
\begin{eqnarray*}
\fl \frac{\rmd}{\rmd t}(\| \Pi \|_h^2 + \| \tilde X \|_h^2)  =\frac{\beta}{6}(\tilde X_{-1}\tilde X_1 + \Pi_{-1}\Pi_1 +\tilde X_{-2}\tilde X_0 + \Pi_{-2}\Pi_0
- 8 \Pi_0 \Pi_{-1}- 8 \tilde X_{-1} \tilde X_0)\\
 +2(\alpha-1)\tilde X_{-1}\Pi_0 - 2 \alpha
\Pi_{-1}\tilde X_0 \le {\rm const} \sum_{j=-2}^1 (|\Pi_j|^2 +|\tilde X_j|^2)\nonumber.
\end{eqnarray*}
As in the second order accurate case, we explicitly solve the Laplace
transformed problem for vanishing initial data and no forcing term and
write the solution as
\begin{eqnarray}
\hat \Pi_j &=& c^{\Pi}_{j\Pi_1} \hat g^{\Pi}_1 + c^{\Pi}_{j\Pi_2} \hat g^{\Pi}_2 + c^{\Pi}_{jX_1} \hat g^{X}_1 + c^{\Pi}_{jX_2} \hat g^{X}_2 , \\
\hat X_j &=& c^X_{j\Pi_1} \hat g^{\Pi}_1 + c^X_{j\Pi_2} \hat g^{\Pi}_2 + c^X_{jX_1} \hat g^{X}_1 + c^X_{jX_2} \hat g^{X}_2.
\end{eqnarray}
We numerically verify the Kreiss condition by plotting the quantity
\begin{equation}
N = \left( \sum_{\tiny\begin{array}{c}
j=-2,\ldots,1\\
k=\Pi_1, \Pi_2, X_1, X_2
\end{array}}( |c^{\Pi}_{jk}|^2 + |c^{X}_{jk}|^2) \right)^{1/2}.
\end{equation}
inside the semi-disk $S$ with $C_0=30$.  

The modified homogeneous boundary conditions for the auxiliary problem are
\begin{eqnarray}
\tilde X_{-1} &=& \frac{1}{\beta}\left(-48 \tilde X_{0}-6 \tilde X_{1}+\frac{72(\alpha-1)}{\beta}  \Pi_{0}\right), \\
\tilde X_{-2} &=& -\frac{6}{\beta^3} (65 \beta^2+144 \alpha^2) \tilde X_{0}, \\
\Pi_{-1} &=&
\frac{1}{\beta}\left(48 \frac{2-\alpha}{\alpha} \Pi_{0}-6 \Pi_{1}-\frac{72 \alpha}{\beta} \tilde X_{0}\right), \\
\Pi_{-2} &=&
-\frac{6}{\beta^3 \alpha} ((65 \alpha -128) \beta^2+12(\alpha-1)) \Pi_{0}-\frac{96}{\beta \alpha} \Pi_{1},
\end{eqnarray}
and they give the estimate
\begin{eqnarray}
\fl\frac{\rmd}{\rmd t}(\| \Pi \|_h^2 + \| \tilde X \|_h^2) \le - \sum_{j=0}^1
(|\Pi_j|^2 + |\tilde X_j|^2)+\| \Pi \|_h^2 + \| \tilde X \|_h^2 + \| F^{\Pi}\|_h^2 + \| F^X
  \|_h^2.
\end{eqnarray}
This implies inequalities (\ref{Eq:normwestimate}) and
(\ref{Eq:wjestimate}) and strong stability with respect to the $D_+$
norm follows from the fact that
$\| \Pi\|_h^2 + \| \tilde X \|^2_h$ is equivalent to $\|\Pi\|_h^2 + \|
D_+\phi\|^2_h$.  Convergence is a consequence of estimates for
the error.  However, due to the modified forcing term (\ref{Eq:FX2})
we are only able to show that $\| \Pi\|_h^2 + \| \tilde X\|_h^2 <
\Or(h^7)$.  Given that the numerical tests of Section~\ref{Sec:Numerical}
indicate fourth order convergence, we believe that this estimate is
not optimal.

\subsubsection{Time-like boundary}
\label{Sec:timelike4}

In the time-like case ($0\neq|\beta| < 1$) the discrete boundary
conditions are
\begin{eqnarray}
\Pi_0 - D^{(1)} \phi_0 &=& g^{\phi}_1 \label{Eq:shift4TL1g}, \\
h^5 D_+^5\phi_{-2} &=& g^{\phi}_2\label{Eq:shift4TL2g}, \\
h^4D_+^4\Pi_{-1} &=& g^{\Pi}_1\label{Eq:shift4TL3g}, \\
h^4D_+^4\Pi_{-2} &=& g^{\Pi}_2\label{Eq:shift4TL4g},
\end{eqnarray}
where $D^{(1)}$ is given in (\ref{Eq:D1}).  We introduce $\tilde X_j =
\tilde D_+ \phi_j$ for $j\ge 0$, where $\phi_{-1}$ in $\tilde X_0$ is
given by system (\ref{Eq:shift4TL1g}-\ref{Eq:shift4TL2g}) with
$g^{\phi}_k = 0$.  Note that $\tilde X_0$ contains also $\Pi_0$.  A
discrete reduction to first order gives the ghost-point
prescriptions
\begin{eqnarray}
\fl \Pi_{-1} = 4 \Pi_{0}-6 \Pi_{1}+4
\Pi_{2}-\Pi_{3}+g^{\Pi}_1 - \frac{4}{\beta} g^X_1 + \frac{1}{3\beta}g^X_2,\label{Eq:gz4TL1} \\
\fl \Pi_{-2} = 10 \Pi_{0}-20 \Pi_{1}+15 \Pi_{2}-4 \Pi_{3}+4
g^{\Pi}_1+g^{\Pi}_2 - 32 \frac{16\alpha-17}{\beta(10\alpha-9)}g^X_1\label{Eq:gz4TL2} \\+
\frac{8}{3} \frac{16\alpha-17}{\beta(10\alpha-9)}g^X_2,\nonumber \\
\fl \tilde X_{-1} =
-\frac{6 \alpha-19}{10 \alpha-9} \tilde X_{0}+\frac{18 \alpha-5}{10
  \alpha-9} \tilde X_{1}-\frac{2 \alpha-1}{10 \alpha-9} \tilde
X_{2}-\frac{24}{10\alpha-9} \Pi_{0}-\frac{2}{3}\frac{30\alpha-79}{10\alpha-9} g^{X}_1\label{Eq:gz4TL3} \\
+\frac{8}{9}\frac{3 \alpha-4}{10\alpha-9} g^{X}_2, \nonumber \\ 
\fl \tilde X_{-2} = -\frac{-383 \beta-1556+1448 \alpha+358 \beta
  \alpha}{\beta (118\alpha-127)} \tilde X_{0}+\frac{2}{3} \frac{528
  \alpha-552+866 \beta \alpha-1001 \beta}{\beta (118\alpha-127)}
\tilde X_{1}\label{Eq:gz4TL4}\\-\frac{1}{3} \frac{690 \beta \alpha+168 \alpha-180-773
  \beta}{\beta (118\alpha-127)} \tilde X_{2}+\frac{2}{3} \frac{54 \alpha-59}{118\alpha-127} \tilde
X_{3}-\frac{1}{3} \frac{10\alpha-9}{118\alpha-127} \tilde
X_{4}\nonumber\\+\frac{4}{3} \frac{666 \beta \alpha-287-697 \beta+270
  \alpha}{\beta (118\alpha-127)} \Pi_{0}-24 \frac{59+54 \beta
  \alpha-54 \alpha-59 \beta}{\beta (118\alpha-127)} \Pi_{1}\nonumber \\
+12 \frac{59+54 \beta \alpha-54 \alpha-59 \beta}{\beta (-127+118
  \alpha)} \Pi_{2}-\frac{8}{3} \frac{59+54 \beta \alpha-54 \alpha-59
  \beta}{\beta (118\alpha-127)} \Pi_{3}\nonumber \\
-\frac{2}{3}
\frac{708+2376\alpha\beta-2596\beta-4051\beta^2-648\alpha+3750\alpha\beta^2}{\beta^2
  (118\alpha-127)} g^{X}_1\nonumber\\
+\frac{-648\alpha+1728\alpha\beta-5865\beta^2+708-1888\beta+5442\alpha\beta^2}{18\beta^2 
  (118\alpha-127)} g^{X}_2\nonumber\\
+\frac{2}{3} \frac{-162 \alpha+216 \beta
  \alpha-236 \beta+177}{\beta (118\alpha-127)} g^{\Pi}_1-\frac{2}{3}
\frac{54 \alpha-59}{118\alpha-127} g^{\Pi}_2, \nonumber
\end{eqnarray}
where $g^X_1 = g^{\phi}_1$ and $g^X_2 = g^{\phi}_2/h$.  The forcing
terms near the boundary are
\begin{eqnarray*}
F^X_2 &=& \tilde D_+ F^{\phi}_2  +
 \frac{\alpha\beta}{36h}(g^{X}_2 -12g^{X}_1), \\
F^X_1 &=& \tilde D_+ F^{\phi}_1,
\end{eqnarray*}
and $F^X_0$ is obtained from the definition of $\tilde X_0$ with the
replacements $\phi_j \to F^{\phi}_j$, $\Pi_j \to F^{\Pi}_j$.
 
The rest of the proof proceeds as in the outflow case.  
That the scheme is convergent follows from estimates
for the errors.  As in the outflow case, we are only able to show 3.5th
order convergence, although experiments suggest that the scheme is
fourth order convergent.

\section{The fully discrete initial-boundary value problem: numerical tests}
\label{Sec:Numerical}

The semi-discrete schemes we have considered here can be turned into
fully discrete schemes using, for example, fourth order Runge-Kutta as
a time integrator.  In general, stability of the semi-discrete scheme
does not guarantee that the fully discrete scheme is stable
\cite{LT,T}.  We therefore check stability of the fully discrete
scheme by performing numerical convergence tests.

The convergence tests are performed using the domain $0 \le x \le L$,
where $x=0$ is the physical boundary, and by monitoring the errors in
the interval $0\le x\le 1$ at time $t=1$.  The value of $L$ is chosen large
enough so that the numerical solution in the interval $0\le x\le 1$
for $t\le 1$ is
numerically unaffected by the boundary at $x=L$. We have used a
Courant factor of $0.5$ and resolutions ranging from $h=1/25$
to $1/400$. The code is tested against the exact solution $\phi(x,t) =
f(-x+(1-\beta)t)$, $\Pi(x,t) = f'(-x+(1-\beta)t)$, where $f(x) =
\sin(2\pi x)$. 

\begin{table}[h]
\caption{The discrete boundary conditions described in Section
  \ref{Sec:Summary} give the expected order of convergence.}
\begin{center}
\begin{tabular}{c@{\qquad}cc}
\multicolumn{3}{c}{Errors and convergence rates}\\
\hline
$N$ & $l_2$ & $q$\\
\hline
\multicolumn{3}{l}{(a) Second order accurate case, $\beta = 2$}\\
$25$ & $7.35084\,10^{-1}$ & \\
$50$ & $1.83951\,10^{-1}$ & $1.9986$\\
$100$ & $4.60081\,10^{-2}$ & $1.9994$\\
$200$ & $1.15021\,10^{-2}$ & $2.0000$\\
$400$ & $2.87555\,10^{-3}$ & $2.0000$\\
\hline
\multicolumn{3}{l}{(b) Second order accurate case, $\beta = -1/5$}\\
$25$ & $1.06042\,10^{-1}$ & \\
$50$ & $2.59231\,10^{-2}$ & $2.0323$\\
$100$ & $6.41559\,10^{-3}$ & $2.0146$\\
$200$ & $1.59673\,10^{-3}$ & $2.0065$\\
$400$ & $3.98366\,10^{-4}$ & $2.0030$\\
\hline
\multicolumn{3}{l}{(c) Fourth order accurate case, $\beta = 2$}\\
$25$ & $9.70747\,10^{-3}$ & \\
$50$ & $6.10334\,10^{-4}$ & $3.9914$\\
$100$ & $3.82024\,10^{-5}$ & $3.9979$\\
$200$ & $2.38809\,10^{-6}$ & $3.9997$\\
$400$ & $1.49255\,10^{-7}$ & $4.0000$\\
\hline
\multicolumn{3}{l}{(d) Fourth order accurate case, $\beta = -1/5$}\\
$25$ & $1.01955\,10^{-3}$ & \\
$50$ & $6.71790\,10^{-5}$ & $3.9238$\\
$100$ & $4.30128\,10^{-6}$ & $3.9652$\\
$200$ & $2.72064\,10^{-7}$ & $3.9827$\\
$400$ & $1.71025\,10^{-8}$ & $3.9917$\\
\hline
\end{tabular}
\label{Tab:convergence}
\end{center}
\end{table}

As shown in table \ref{Tab:convergence}, we find good second or
fourth-order convergence in the norm $(\| \Pi\|_h^2 + \|
D_+\phi\|_h^2)^{1/2}$ over all resolutions. We focused on the $\beta =
2$ and $\beta = -1/5$ cases, but the schemes are convergent for other
values of the shift parameter in the ranges $\beta>1$ and $|\beta|
<1$. However, we noticed that in the time-like, fourth order accurate
case, in order to avoid losing accuracy, the boundary data $g$ had to
be Taylor expanded at the intermediate time steps of the Runge-Kutta
integrator.  This issue is discussed in \cite{Joh}.

\section{Application to the Nagy-Ortiz-Reula formulation of the
  Einstein equations in spherical symmetry} 
\label{Sec:NOR}

The main interest of the one dimensional shifted wave equation is as a
toy model for second order in space formulations of three-dimensional
general relativistic initial-boundary value problems.  For such
problems initial and boundary data cannot be freely specified.  The
initial data has to satisfy the constraints and the boundary
conditions should be such that no constraint violation is injected.
Whereas maximally dissipative boundary conditions involve only first
derivatives across the boundary, constraint-preserving boundary
conditions require second derivatives across the boundary.  

We consider the Nagy-Ortiz-Reula (NOR) formulation \cite{NOR} of the
Einstein equations, restricted to spherical symmetry and coupled to a
massless minimally coupled scalar field.  In spherical symmetry the
line element takes the form
\begin{equation}
ds^2 = -\alpha^2 dt^2 + g_{rr} (dr + \beta dt)^2 + g_{\theta\theta}
(d\theta^2 + \sin^2 \theta d\phi^2).
\end{equation}
and the reduction of the NOR system to spherical symmetry is
straightforward.  However, if the origin is included in the domain,
the definition of the auxiliary variable $f$ and the addition of its
definition constraint to the right hand side of the evolution
equations require consideration. We use the
regularised dynamical variables $g_T\equiv r^{-2} g_{\theta\theta}$,
$K_T\equiv r^{-2} K_{\theta\theta}$, and $f\equiv - 2(\ln g_T)_{,r}$,
and instead of adding a multiple of ${\cal G}_{,r}$ to the right-hand
side of the evolution equation for $K_{rr}$, where ${\cal G}\equiv
f+2(\ln g_T)_{,r}$ is the definition constraint for the auxiliary
variable $f$, we add ${\cal G}_{,r}-{\cal G}/r$.  The regularity
conditions are the origin are that $g_{rr}$, $g_T$, $K_{rr}$, $K_T$,
$\phi$, $\Pi$ are even functions of $r$ satisfying $g_T|_{r=0} =
g_{rr}|_{r=0}$, $K_T|_{r=0} = K_{rr}|_{r=0}$ and $f$ is an odd
function of $r$.

With fixed densitised lapse, $\alpha = \sqrt{g_{rr}}g_T Q$, and fixed
shift, the evolution equations are
\begin{eqnarray}
\fl \p_t g_{rr} &=& \beta \p_r g_{rr}+2 g_{rr} \p_r \beta-2 \alpha K_{rr}, \label{Eq:ADM1}\\
\fl \p_t g_T &=& \beta \p_r g_T+\frac{2\beta}{r} g_T-2 \alpha K_T,
\label{Eq:ADM2}\\
\fl \p_t K_{rr} &=& \beta \p_r K_{rr}-\frac{\alpha}{2g_{rr}} \p^2_r
g_{rr}+\alpha \p_r f + \frac{\alpha}{2g_{rr}^2} (\p_r g_{rr})^2+2
K_{rr} \p_r \beta-\frac{2 \alpha}{g_TQ} \p_r g_T \p_r,
Q\\
\fl &&-\frac{3\alpha}{2g_T^2} (\p_r g_T)^2-\frac{\alpha}{g_{rr}}
K_{rr}^2-\frac{\alpha}{Q} \p^2_r Q-\frac{\alpha}{r}
f-\frac{4\alpha}{rg_T} 
\p_r g_T+\frac{\alpha}{rg_{rr}} \p_r g_{rr}\nonumber\\
\fl &&+\frac{2\alpha}{g_T} K_T,
K_{rr}-\frac{\alpha}{2g_{rr}Q} \p_r g_{rr} \p_r Q -8 \pi \alpha (\p_r
\phi)^2, \nonumber\\
\fl \p_t K_T &=& \beta \p_r K_T-\frac{\alpha}{2g_{rr}} \p_r^2
g_T + \frac{2\beta}{r} K_T-\frac{3\alpha}{rg_{rr}}\p_r
g_T+\frac{\alpha}{r^2}\left(1-\frac{g_T}{g_{rr}}\right)
+\frac{\alpha}{g_{rr}} K_{rr} K_T \\
\fl && -\frac{\alpha}{2g_{rr}g_T} (\p_r g_T)^2-\frac{\alpha}{2g_{rr}Q}
\p_r g_T\p_r Q-\frac{\alpha}{rg_{rr}Q} g_T \p_r Q, \nonumber\\
\fl \p_t f &=& \beta \p_r f+f\p_r\beta+ \frac{4\beta}{r^2}-\frac{4}{r} \p_r
\beta+\frac{2\alpha}{g_Tg_{rr}} K_T \p_r
g_{rr}+\frac{4\alpha}{g_TQ}K_T \p_r Q-\frac{4\alpha}{rg_T}K_T\\
\fl &&+\frac{2\alpha}{g_Tg_{rr}} K_{rr},
\p_rg_T+\frac{4\alpha}{rg_{rr}}K_{rr}+\frac{2\alpha}{g_T^2} \p_r g_T
K_T-16 \alpha \pi \Pi \p_r \phi,\nonumber\\
\fl \p_t \phi &=& \beta \p_r \phi-\alpha \Pi,\\
\fl 
\p_t \Pi &=& \beta \p_r \Pi-\frac{\alpha}{g_{rr}} \p^2_r
\phi-\frac{2\alpha}{g_{rr}g_T} \p_r
g_T \p_r \phi - \frac{2\alpha}{rg_{rr}} \p_r \phi-\frac{\alpha}{Qg_{rr}} 
\p_r Q\p_r \phi + \alpha K \Pi, \label{Eq:SF2}
\end{eqnarray}
where $K = K_{rr}/g_{rr} + 2K_T/g_T$.

The Hamiltonian and momentum constraints and the constraint defining
$f$ are
\begin{eqnarray}
\fl {\cal C} &\equiv& -\frac{\p_r^2 g_T}{g_{rr}g_T} + \frac{\p_r g_{rr}
  \p_r g_T}{2g_{rr}^2g_T} +\frac{1}{r^2g_T}-\frac{1}{r^2g_{rr}} +
  \frac{\p_r g_{rr}}{rg_{rr}^2} - \frac{3\p_r g_T}{rg_{rr}g_T} +
  \frac{K_T^2}{g_T^2} + \frac{2K_{rr}K_T}{g_{rr}g_T} \\
\fl && + \frac{(\p_r
  g_T)^2}{4g_{rr}g_T^2}-4\pi\left(\Pi^2 +\frac{(\p_r\phi)^2}{g_{rr}}\right)=0,\nonumber\\
\fl {\cal C}_r &\equiv& -2\frac{\p_r K_T}{g_T} + \left( \frac{K_T}{g_T^2}
+ \frac{K_{rr}}{g_Tg_{rr}}\right) \p_r g_T +
\frac{2}{r}\left(\frac{K_{rr}}{g_{rr}} - \frac{K_T}{g_T}\right) -8\pi\Pi\p_r\phi=0,\\
\fl {\cal G} &\equiv& f +\frac{2\p_r g_T}{g_T} = 0. \label{Eq:G}
\end{eqnarray}

System (\ref{Eq:ADM1})--(\ref{Eq:SF2}) is symmetric hyperbolic with a
symmetric hyperbolic constraint evolution system.  The characteristic
speeds and variables are
\begin{eqnarray}
\beta, &\qquad& w^0 = f,\\
\beta \pm \frac{\alpha}{\sqrt{g_{rr}}}, &\qquad& w^{\pm}_T = \p_r g_T
\mp 2\sqrt{g_{rr}}K_T, \\
\beta \pm \frac{\alpha}{\sqrt{g_{rr}}}, &\qquad&
w^{\pm}_{rr}= \p_r g_{rr}
\mp 2\sqrt{g_{rr}}K_{rr} -2g_{rr}f,\\
\beta \pm \frac{\alpha}{\sqrt{g_{rr}}},&\qquad& w^{\pm}_{\phi} = \p_r \phi \mp \sqrt{g_{rr}}\Pi.
\end{eqnarray}
The characteristic constraint variables are
\begin{eqnarray}
{\cal C}^0 = {\cal G},\\
{\cal C}^{\pm} = {\cal C}_r \mp \sqrt{g_{rr}} {\cal C} 
- \left(\frac{2K_T}{g_T} \mp \frac{\p_r g_T}{\sqrt{g_{rr}}g_T} \mp
\frac{2}{r\sqrt{g_{rr}}}\right){\cal G}.
\end{eqnarray}
Note that in spherical symmetry, some of the highest derivative terms
that are present generically in 3D cancel, both in the main evolution
equations and in the implied constraint evolution. In particular, the
right-hand side of $\dot {\cal C}$ and $\dot {\cal C}_i$ does not
contain ${\cal G}_{i,jk}$ terms. For this reason $\cal G$ appears
undifferentiated in ${\cal C}^{\pm}$, whereas in 3D its first
derivatives would appear.

If we introduce an artificial outer boundary and assume that
$0<\beta<\alpha/\sqrt{g_{rr}}$ at the outer boundary, we have four
incoming characteristic variables for the main system and two incoming
characteristic constraints.  The constraint preserving boundary
conditions are
\begin{eqnarray}
&&w^{+}_{rr} = g^{+}_{rr},\\
&&w^{+}_{\phi} = g^{+}_{\phi},\\
&&{\cal G} = 0,\\
&&{\cal C}_r - \sqrt{g_{rr}} {\cal C}  = 0. \label{Eq:Cplus}
\end{eqnarray}
where $g^+_{rr}$ and $g^+_{\phi}$ are two freely specifiable functions
compatible with the initial data.  

We generalize the discrete boundary conditions introduced in the first
part of the paper in the following way.  We extrapolate all fields
using fifth order extrapolation and populate $g_{rr}$, and $\phi$ at
the two ghost points, $r_{N+1}$ and $r_{N+2}$, by solving
\begin{eqnarray}
&&D_0 \left(1-\frac{h^2}{6}D_+D_-\right)g_{rr} - 2\sqrt{g_{rr}}K_{rr} -2g_{rr}f = g^{+}_{rr},\label{Eq:cpbc2}\\
&&h^6 D_-^6g_{rr}|_{N+2} = 0,\\
&&D_0 \left(1-\frac{h^2}{6}D_+D_-\right)\phi - \sqrt{g_{rr}}\Pi = g^{+}_{\phi},\label{Eq:cpbc3}\\
&&h^6 D_-^6\phi|_{N+2} = 0,
\end{eqnarray}
where (\ref{Eq:cpbc2}) and (\ref{Eq:cpbc3}) are evaluated at $r_N$.
We then solve (\ref{Eq:Cplus}) for $\p_r^2 g_{T}$ and use its fourth
order accurate approximation, combined with sixth order extrapolation
at $r_{N+2}$, to populate $g_T$ at the ghost zones (we are able to
compute the first derivative of $g_T$ from the extrapolation and the
first derivatives of $g_{rr}$ and $\phi$ from the two maximally
dissipative boundary conditions above).  Finally, we impose ${\cal G}
= 0$ by giving data to $f$ at $r_N$ using
\begin{equation}
f = -\frac{2}{g_T}D_0  \left(1-\frac{h^2}{6}D_+D_-\right) g_T.
\end{equation}

We evolve Schwarzschild space-time in standard Kerr-Schild
coordinates, obtaining the initial data and the fixed lapse and shift
from the exact solution. We introduce a space-like (outflow) boundary
to excise the singularity.  At this boundary we use fifth order
extrapolation for the extrinsic curvature, $f$ and $\Pi$ and sixth
order for the 3-metric components and $\phi$. In order to obtain a
non-vacuum solution, we inject a
scalar field pulse through the outer boundary by using 
\begin{equation}
g_{\phi}^+ = A\sin^8(t-t_0), \qquad t_0 < t < t_0+\pi
\end{equation}
where $A= 0.02$ and $t_0 = 0.1$, and monitor the $L_2$-norm of the
Hamiltonian constraint at times $t=8M$, after the pulse has completely
entered the domain, and $t=16M$, after the pulse has fallen into the
black hole.  See table \ref{Tab:convergence2}.  We obtain similar
convergence rates for ${\cal C}_r$ and ${\cal G}$.

The domain extends from $r = 1.8M$ and $r = 11.8M$.  We set $M=1$, use
a Courant factor of $0.1$ and a dissipation parameter $\sigma = 0.05$.
Dissipation is switched off at the last few grid points near the outer
boundary, as we have observed a numerical instability when
Kreiss-Oliger dissipation is used in combination with fifth and sixth
order extrapolation.  On the other hand, according to our tests a
lower order of extrapolation only gives third order convergence for
the contraints.

\begin{table}[h]
\caption{We give initial data corresponding to a Schwarzschild black
  hole in Kerr-Schild coordinates, inject a scalar field pulse through
  the outer boundary and monitor the constraints.  Here we give the
  discrete $L_2$-norm and convergence rate for the Hamiltonian
  constraint.}
\begin{center}
\begin{tabular}{c@{\qquad}cc}
\multicolumn{3}{c}{Errors and convergence rates}\\
\hline
$N$ & $l_2$ & $q$\\
\hline
\multicolumn{3}{l}{(a) $t=8M$}\\
$100$ & $3.68824\,10^{-4}$ & $3.6358$\\
$200$ & $2.96704\,10^{-5}$ & $4.4761$\\
$400$ & $1.33308\,10^{-6}$ & $4.3550$\\
$800$ & $6.51396\,10^{-8}$ &         \\
\hline
\multicolumn{3}{l}{(b) $t=16M$}\\
$100$ & $9.88525\,10^{-5}$ & $3.9867$\\
$200$ & $6.23527\,10^{-6}$ & $4.3909$\\
$400$ & $2.97205\,10^{-7}$ & $4.4295$\\
$800$ & $1.37924\,10^{-8}$ &         \\
\hline
\end{tabular}
\label{Tab:convergence2}
\end{center}
\end{table}

\section{Conclusions}

We have constructed second and fourth order accurate approximations of
maximally dissipative boundary conditions for the shifted wave
equation and have shown strong stability of the semi-discrete scheme
using the Laplace transform method.  Two important steps were
obtaining an appropriate discrete reduction to first order,
and verifying that the solution of the problem with trivial initial
data and no forcing term can be estimated in terms of the boundary
data (the Kreiss condition). The Kreiss condition was verified by
plotting the function $N(\tilde s)$ to show that it is bounded.  We
have also shown numerically that the fully discrete schemes obtained
by integrating our semi-discrete systems with a fourth-order
Runge-Kutta time integrator are second or fourth-order accurate.

Whereas semi-discrete approximations of one-dimensional first-order
hyperbolic systems with constant coefficients can be transformed by a
change of variables into a set of advection equations, one for each
characteristic variable, the same is not true for second-order
systems\footnote{As an example consider the system $\rmd \phi_j /\rmd
t= \Pi_j$, $\rmd \Pi_j/\rmd t = D_+D_- \phi_j$, whose characteristic
variables in Fourier space are $\hat \Pi \pm 2\rmi/h \sin (\xi/2) \hat
\phi$.  It is not possible to translate these variables back into
physical space.}. Therefore numerical methods for second-order in
space systems need to be developed independently. Our results for the
second-order in space wave equation suggest a general heuristic
prescription for discretising boundary conditions in general
second-order in space hyperbolic systems: {\em Populate all necessary
ghost zones by extrapolation, combined with centered difference
approximations of the continuum boundary conditions at the boundary
point.}  We have given a simple example, the NOR system with scalar
field matter in spherical symmetry, where this prescription gives a
stable fourth-order accurate scheme.

The interior schemes discussed in this paper always use a minimal
width centered discretization, even in the outflow case ($\beta > 1$).
Despite reports on the benefits of the upwind (one-sided)
discretization of the shift terms \cite{AB,YBS}, we find that, at
least in the linear constant coefficient case, this is not
necessary. Note that in contrast to an upwind scheme our
semi-discrete central scheme is non-dissipative, as the Cauchy problem
(without boundaries) admits a conserved energy.

\ack

We wish to thank Ian Hawke, Ian Hinder, Heinz-Otto Kreiss, Luis Lehner
and Olivier Sarbach for helpful discussions, suggestions and/or
comments on the manuscript.  This research was supported by a
Marie Curie Intra-European Fellowship to GC within the 6th European
Community Framework Program.

\appendix

\section{The discrete energy method and summation by parts}
\label{Sec:DEM}

To derive continuum energy estimates one uses integration by parts to
generate boundary terms which can be controlled by imposing suitable
boundary conditions.  To obtain similar estimate for the semi-discrete
problem one can use difference operators satisfying the {\it summation
by parts} rule \cite{Strand}.  Discretisations of $\rmd^2/\rmd x^2$ with that
property have been constructed in \cite{MatNor}. 

In this Appendix, we attempt to use summation by parts methods to
construct stable (and sufficiently accurate) schemes for the shifted
wave equation. We are not successful, and try to explain why.

We adopt the notation of reference \cite{MatNor} and
definitions used there, and write grid functions as column vectors, so
that for two grid functions $u_j$ and $v_j$
\begin{equation}
(u,v)\equiv \sum_{j=0}^N u_j v_j \equiv u^{\rm T} v.
\end{equation}
In contrast to the body of this paper, we assume that there are two
boundaries in $x$, but the case with only one boundary can be obtained
trivially by setting $N\to\infty$ and ignoring all terms arising at
the right boundary in what follows.  A more general inner product on
grid functions can be characterised by a positive definite symmetric
$(N+1)\times (N+1)$ matrix $H$,
\begin{equation}
(u,v)_H\equiv  u^{\rm T} H v, \quad H^{\rm T}=H>0.
\end{equation}
Its continuum limit should be the $L_2$ inner product. In a similar
way, discrete derivative operators can be written as matrices. 

In this notation, the semi-discrete shifted wave equation can be
written as
\begin{eqnarray*}
\frac{\rmd}{\rmd t}\phi&=&\beta \hat D_1\phi+\Pi, \\
\frac{\rmd}{\rmd t}\Pi&=&\beta D_1\Pi+D_2\phi,
\end{eqnarray*}
where $\hat D_1$ and $D_1$ are approximations to $\rmd/\rmd x$ and $D_2$ is
an approximation to $\rmd^2/\rmd x^2$. Similarly, we can write the discrete
energy as
\begin{equation}
E=\Pi^{\rm T} H\Pi+\phi^{\rm T}A\phi. \label{Eq:EHA}
\end{equation}
where $A^{\rm T} = A$ and $\phi^T A\phi$ represents
$\int(\p_x\phi)^2\rmd x$. In particular, it should have the positivity properties
\begin{eqnarray}
\label{energy1}
\phi^{\rm T}A\phi &\ge & 0 \hbox{ for all } \phi, \\
\label{energy2}
\phi^{\rm T}A\phi &=&0 \hbox{ if and only if } D_+\phi_j=0
\end{eqnarray}

We assume $\beta>1$ and look for (sufficiently accurate) difference
operators and matrices $H$ and $A$ that give a discrete energy
estimate.  Taking a time derivative of the discrete energy
(\ref{Eq:EHA}), we have
\begin{eqnarray}
\frac{\rmd}{\rmd t} E &=& 2\beta(\Pi^{\rm T} H D_1 \Pi + \phi^{\rm T}A \hat D_1
\phi) \\
&& + 2\Pi^{\rm T} H D_2 \phi + 2\Pi^{\rm T} A \phi \nonumber.
\end{eqnarray}
The key point is that the requirement that this expression be a pure
boundary term leads to only one summation by parts condition for
$\beta=0$, but to three separate conditions for $\beta\ne 0$, namely 
\begin{eqnarray}
2 \Pi^{\rm T} H D_1 \Pi &=& \Pi^2|_0^N\label{Eq:SBP1}, \\
\Pi^{\rm T} H D_2 \phi &=& - \Pi^{\rm T} A \phi + (\Pi S \phi)|_0^N\label{Eq:SBP2}, \\
2\phi^{\rm T} A \hat D_1 \phi &=& (S\phi)^2|_0^N\label{Eq:SBP3}
\end{eqnarray}
where $S\phi$ is an approximation of $\p_x\phi$ at the boundaries.

We now assume that (in the terminology of \cite{MatNor}) $D_1$ is a
{\em first derivative summation by parts operator}, meaning that
\begin{equation}
\label{SBP1}
HD_1+(HD_1)^{\rm T} = B,
\end{equation}
where $B\equiv {\rm diag}(-1,0,\ldots,0,1)$.  We also assume that
$D_2$ is a {\em symmetric second derivative summation by parts
operator}, i.e., it satisfies
\begin{equation}
\label{SBP2S}
HD_2=(-A+BS).
\end{equation}

These two assumptions guarantee that (\ref{Eq:SBP1}) and
(\ref{Eq:SBP2}) are satisfied.  The third condition,
equation \eref{Eq:SBP3}, gives 
\begin{equation}
\label{cond1a}
A\hat D_1+(A\hat D_1)^{\rm T} = S^{\rm T} B S + M,
\end{equation}
where $M=M^{\rm T}\le 0$.  

Assuming a minimal width second order accurate approximation in the
interior, we have that \cite{MatNor}
\begin{eqnarray}
H&=&h\,{\rm diag}(1/2,1,\ldots,1,1/2), \\
D_1&=&(D_+,D_0,\ldots,D_0,D_-), \\
D_2&=&(D_+^2,D_+D_-,\ldots,D_+D_-,D_-^2), \\
BS&=&(D_+-h/2\,D_+^2,0,\ldots,0,D_--h/2\,D_-^2),
\end{eqnarray}
and $\phi^{\rm T} A \phi = \sum_{j=0}^{N-1} (D_+\phi_j)^2 h$.  We see
that the matrix $A$ satisfies the desired properties.  We need to
construct an operator $\hat D_1$ approximating $\p_x$ which coincides
with $D_0$ at the interior.  We consider 
\begin{eqnarray}
(\hat D_1 \phi)_0 &=& aD_+\phi_0 + (1-a) D_+\phi_1, \\
(\hat D_1 \phi)_1 &=& bD_+\phi_0 + (1-b) D_+\phi_1
\end{eqnarray}
and find that if $2a=8b=3$, condition (\ref{Eq:SBP3}) is satisfied.
However, this approximation is not accurate enough: it is only first
order convergent.

Alternatively, one could consider the modification
\begin{eqnarray}
(\hat D_1 \phi)_0 &=& \frac{3}{2}D_+\phi_0  -\frac{1}{2}D_+\phi_1,\\
(\hat D_1 \phi)_1 &=& D_0\phi_1 + ah^2 D_+^3 \phi_0.
\end{eqnarray}
We find that $M$ in equation \eref{cond1a} is indefinite (the product of
two of its eigenvalues is negative) unless $a=0$, in which case it is
semi-definite positive.

Clearly there is an infinity of other choices to make, and it is
difficult to exhaust all possibilities.  We have made a number of
attempts aimed at obtaining summation by parts for the shifted wave
equation, but we have not been able to construct operators and scalar
products which give second order convergent schemes.  We also note
that a direct use of the operators of Appendix C.1 of \cite{MatNor},
i.e., using $D_+$ and $D_+^2$ to approximate first and second
derivatives at the boundary, gives rise to a first order reflection
from the boundary.  With the same settings used in Section
\ref{Sec:Numerical}, we carried out convergence tests for $\beta =2$ to
confirm this.  See table \ref{Tab:convergence3}.

\begin{table}[h]
\caption{Using $D_+$ and $D_+^2$ to approximate first and second
  derivatives at the outflow boundary gives only first order accuracy.  Here
  we used $\beta = 2$.}
\begin{center}
\begin{tabular}{c@{\qquad}cc}
\multicolumn{3}{c}{Errors and convergence rates}\\
\hline
$N$ & $l_2$ & $q$\\
\hline
\multicolumn{3}{l}{}\\
$50$  & $6.54655\,10^{-1}$ & $1.0630$\\
$100$ & $3.13340\,10^{-1}$ & $1.0252$\\
$200$ & $1.53950\,10^{-1}$ & $1.0065$\\
$400$ & $7.66263\,10^{-2}$ & \\
\hline
\end{tabular}
\label{Tab:convergence3}
\end{center}
\end{table}

\section{The advection equation}
\label{Sec:advection}

In this Appendix we list a number of known results regarding the
discrete boundary treatment for the advection equation and give a
simple prescription for the inflow fourth order accurate case.

\subsection{Second order accuracy}

The semi-discrete initial-boundary value problem with an outflow
boundary, 
\begin{eqnarray}
&&\frac{\rmd}{\rmd t}v_j = a D_0v_j\qquad j=0,1,2,\ldots \label{Eq:adv2}, \\
&&h^qD_+^qv_{-1} = 0 \label{Eq:adv2_extrap}
\end{eqnarray}
with $a>0$ is strongly stable and second order convergent for $q\ge
2$. For an inflow boundary, $a<0$, the scheme
\begin{eqnarray}
&&\frac{\rmd}{\rmd t}v_j = a D_0v_j\qquad j=1,2,\ldots\\
&&v_{0} = g
\end{eqnarray}
is strongly stable and second order convergent. 

However, this last scheme is actually strongly stable for any $a$,
even if we impose data on an outflow boundary. Moreover, if the
redundant boundary data is second order accurate with respect to the
continuum solution, the scheme is also second
order convergent.

\subsection{Fourth order accuracy}

As shown in \cite{GKO} the semi-discrete initial-boundary
value problem
\begin{eqnarray}
&&\frac{\rmd}{\rmd t}v_j = a D_0\left( 1 - \frac{h^2}{6}D_+D_-\right)
v_j\qquad j=0,1,2,\ldots \label{Eq:adv1}, \\
&&h^qD_+^qv_{-1} = h^qD_+^qv_{-2} = 0 \label{Eq:adv_extrap}
\end{eqnarray}
with $q\ge 4$ and $a>0$ is strongly stable and fourth order convergent.

In the inflow case ($a<0$) the boundary conditions
\begin{eqnarray}
&&v_0 = g\label{Eq:adv_inflow1}, \\
&&h^qD_+^qv_{-1} = 0\label{Eq:adv_inflow2}
\end{eqnarray}
lead to strong stability for $q = 4$ or $5$ and fourth order
convergence is obtained provided that $g$ is fourth order accurate.
Interestingly, for $q\ge 6$ the
scheme is unstable.  One can show that for $a=-1$, $q=6$, $g=0$, the
scheme admits solutions of the form $\rme^{st} f_j$, where $sh \approx
0.093 \pm 1.308\rmi$.

Note that a set of boundary conditions which lead to strong stability
for any value of $a$ (including the case where boundary conditions are
imposed at an outflow boundary) is
\begin{equation}
v_{-1} = g_{-1}, \qquad v_0 = g_0
\end{equation}
If, in addition, $g_{-1} = u(t,-h) + \Or(h^4)$ and $g_0 = u(t,0) + \Or(h^4)$
the scheme is fourth order convergent \cite{GKO}.

\section{Direct second-order treatment of the shifted wave equation}
\label{Sec:Direct}

Eliminating $\Pi$ from (\ref{Eq:shifted1}-\ref{Eq:shifted2}), we
obtain
\begin{equation}
\label{skw}
\p_t^2\phi=2\beta\p_t\p_x \phi+(1-\beta^2)\p_x^2\phi.
\end{equation}
In \cite{SKW}, this equation is investigated as a toy model for the
Einstein equations in the presence of a shift. The minimal width
second-order accurate discretisation in space is
\begin{equation}
\p_t^2\phi=2\beta\p_t D_0 \phi+(1-\beta^2)D_+D_-\phi,
\end{equation}
but this is stable only for $|\beta|<1$. Therefore in regions where
$|\beta|\ge 1$, the discretisation 
\begin{equation}
\p_t^2\phi=2\beta\p_t D_0 \phi+(D_+D_--\beta^2D_0^2)\phi,
\end{equation}
is used instead, with a blending between the two algorithms in a
transition region. The authors can then construct second-order
accurate and stable boundary treatments for both outflow and time-like
boundaries.  They focus on Dirichlet and Neumann boundary conditions.

The system could be reduced to first order in time, second order in
space form by introducing $\p_t\phi$ as an auxiliary variable. On the
level of a semi-discrete (continuous in time) system this change of
variable is trivial. In particular, the difficulty for $|\beta|\ge 1$
and its resolution would be the same. The crucial difference is not
that (\ref{skw}) is in second-order in time form but that $\Pi$ has
been replaced by $\p_t\phi$ in the equivalent first-order in time,
second-order in space system.

\section*{References}


\end{document}